\documentclass[prb,aps,reprint,nobibnotes,twocolumn,longbibliography]{revtex4-1}

\usepackage{amsmath,amssymb,graphicx,color}
\usepackage{braket}
\usepackage{commath}
\usepackage[colorlinks,
            linkcolor=black,
            anchorcolor=black,
            citecolor=black,
            urlcolor=black,
            ]{hyperref}

\newcommand*{\citen}{}% generate error, if `\citen` is already in use
\DeclareRobustCommand*{\citen}[1]{%
  \begingroup
    \romannumeral-`\x % remove space at the beginning of \setcitestyle
    \setcitestyle{numbers,square}%
    \cite{#1}%
  \endgroup
}

\AtBeginDocument{%
    \newwrite\bibnotes
    \def\bibnotesext{Notes.bib}
    \immediate\openout\bibnotes=\jobname\bibnotesext
    \immediate\write\bibnotes{@CONTROL{REVTEX41Control}}
    \immediate\write\bibnotes{@CONTROL{%
    apsrev41Control,author="08",editor="1",pages="1",title="0",year="1"}}
     \if@filesw
     \immediate\write\@auxout{\string\citation{apsrev41Control}}%
    \fi
}%

\begin{document}

\title{Topological Polarization Singularities in Metaphotonics}

\author{Wenzhe Liu\textsuperscript{1}}
\author{Wei Liu\textsuperscript{2}}
\author{Lei Shi\textsuperscript{1,3}}
\email{lshi@fudan.edu.cn}
\author{Yuri Kivshar\textsuperscript{4}}
\email{yuri.kivshar@anu.edu.au}

\affiliation{$^{1}$State Key Laboratory of Surface Physics, Key Laboratory of Micro- and Nano-Photonic Structures (Ministry of Education) and Department of Physics, Fudan University, Shanghai 200433, China}
\affiliation{$^{2}$College for Advanced Interdisciplinary Studies, National University of Defense Technology, Changsha, Hunan 410073, China}
\affiliation{$^{3}$Collaborative Innovation Center of Advanced Microstructures, Nanjing University, Nanjing 210093, China}
\affiliation{$^{4}$Nonlinear Physics Centre, Research School of Physics, Australian National University, Canberra, ACT 2601, Australia}

\begin{abstract}
Polarization singularities of vectorial electromagnetic fields locate at the positions (such as points, lines, or surfaces) where properties of polarization ellipses are not defined. They are manifested as circular and linear polarizations, for which respectively the semi-major axes and normal vectors of polarization ellipses become indefinite. First observed in conical diffraction in 1830's, the field of polarization singularities has been systematically reshaped and deepened by many pioneers of wave optics. Together with other exotic phenomena such as non-Hermiticity and topology, polarization singularities have been introduced into the vibrant field of nanophotonics, rendering unprecedented flexibilities for manipulations of light-matter interactions at the nanoscale. Here we review the recent results on the generation and observation of {\it polarization singularities in metaphotonics}. We start with the discussion of polarization singularities in the Mie theory, where both electric and magnetic multipoles are explored from perspectives of local and global polarization properties. We then proceed with the discussion of various photonic-crystal structures, for which both near- and far-field patterns manifest diverse polarization singularities characterized by the integer Poincar\'{e} or more general half-integer Hopf indices (topological charges). Next, we review the most recent studies of conversions from polarization to phase singularities in scalar wave optics, demonstrating how bound states in the continuum can be exploited to generate directly optical vortices of various charges. Throughout our paper, we discuss and highlight several fundamental concepts and demonstrate their close connections and special links to metaphotonics. We believe polarization singularities can provide novel perspectives for light-matter manipulation for both fundamental studies and their practical applications.
\end{abstract}

\maketitle

\section{Introduction}

\begin{figure}[htpb]
\includegraphics[scale=.95]{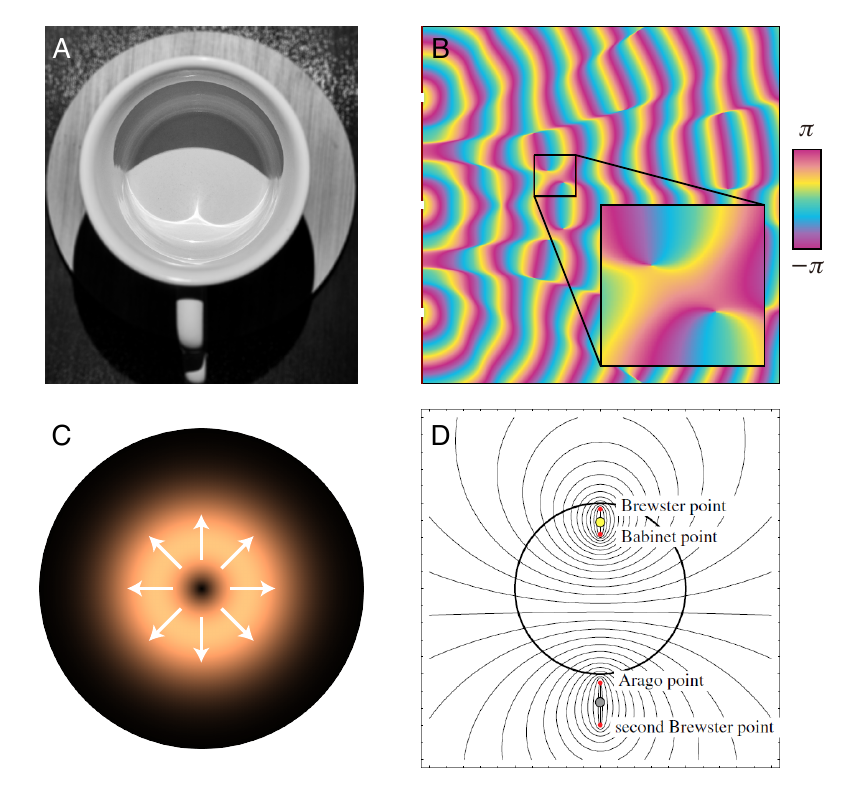}
\caption{Examples of optical singularities. (A) Cusp-shaped caustic in a teacup. (B) Phase singularity in a three-pinhole diffraction. (C) Vector polarization singularity for the TM$_{01}$ waveguide mode. (D) Line-field polarization singularities in clear sky. The image (B) is reproduced from Ref.~\citen{dennis2009singular}, and the image (D) is adapted from Ref.~\cite{berry2004polarization}.}
\label{fig:1}
\end{figure}

Singularities represent a pervasive concept in different areas of mathematics and physics, constituting the fundamental skeletons and frameworks for many related concepts and their applications~\cite{IZUMIYA__Singularity, ARNOLD__Mathematicala, MATSUMOTO_2002__Introduction}.  The field of {\it singular optics} originates from direct observations of singular points in intensity (caustics), phase (vortices), or polarization~\cite{ARNOLD__Singularities,  NYE_natural_1999, berry2001geometry, GBUR_2016__Singular, BERRY_2017__HalfCentury}. At the level of geometric optics, an optical beam is interpreted not as a single ray but as a bunch of rays with the envelopes termed as caustics~\cite{BERRY_1976_Adv.Phys._Waves}. From the perspective of geometric optics, the caustics locate at the positions where the light intensities diverge, and thus become singular. Rainbows are essentially caustics of intensity singularities, as pointed out by R. Potter and G.B. Airy back to 1830s~\cite{potter1838mathematical, airy1838intensity, nussenzveig1977theory}. They also appear as cusp-shaped highlighted curves formed in a teacup when sunlight is reflected by the cups' interior walls (see Fig.~\ref{fig:1}A)~\cite{berry1981singularities,NYE_natural_1999}. However, geometric optics breaks down at these singular points where the intensity divergence is extinguished by wave effects of interference and diffraction.

Singularities are found in many problems of wave optics. At the next level of scalar wave optics, it was realised in the mid-1900s that there are specific regions (could be points or lines) where the phase is ill-defined, and around those regions there are gradient phase distributions and optical power circulations. Those points correspond to phase singularities (optical vortices) \cite{nye1974dislocations, berry1981singularities, berry1998much, dennis2009singular, GBUR_2016__Singular}, a specific example of which is shown in Fig.~\ref{fig:1}B  for the three-slit interference experiment~\cite{wolter1950frage, braunbek1952einzelheiten, born1959principles, masajada2001optical, berry2002exuberant, o2006topology, dennis2009singular}. At the higher level of vectorial wave optics, singularities (generally moving over time) appear where the field vector vanishes at a certain moment. Those singularities are called disclinations~\cite{nye1983polarization}. Interestingly, if a disclination is pinned somewhere, for example at the center of a basic waveguide for the $\rm{TM}_{\rm{01}}$ mode, it could result in a intensity null point (see Fig. \ref{fig:1}C) \cite{snyder2012optical}. Such a stationary disclination is called a vector polarization singularity, which actually was firstly observed in conical diffraction in 1830s and later studied in various anisotropic crystals \cite{hamilton1837third, born1959principles, berry2007conical, jeffrey2007conical, brasselet2009dynamics, fadeyeva2010spatially, leykam2016conical}. As for the steady behaviors of electromagnetic waves, polarization ellipses as the orbits of the oscillating field vectors can be employed for characterizations, which can be represented effectively by line fields in terms of the semi-major or semi-minor axes of the ellipses~\cite{nye1983lines, HOPF_1983__Differential, nye1987wave, NYE_natural_1999, berry2001geometry, berry2004electric}. Elementary singularities of electromagnetic polarization fields are either circularly polarized positions where the semi-major or semi-minor axes are not defined, or linearly polarized positions where the normal vectors perpendicular to the ellipse plane are not defined. Of course, vector polarization singularities are also polarization singularities in such fields, which nevertheless are not generic and can be easily broken into circularly polarized singularities by applying perturbations, provided the polarization charge is an integer. In addition to those fields which are completely polarized, polarization singularities may even appear in a partially polarized field. During the 1800's, researchers discovered polarization singularities in a clear sky where sun light is unpolarized \cite{barral1859oeuvres, babinet1840nouveau, brewster1842existence, brewster1847lxviii, brewster1863xxl, lord1871light, hulst1981light, horvath2002first}, which were then revealed to exhibit interesting polarization patterns~\cite{berry2004polarization} (see Fig.~\ref{fig:1}D).

A special attention has been paid to the study of optical singularities~\cite{NYE_natural_1999, soskin2001singular, dennis2009singular, rubinsztein2016roadmap, GBUR_2016__Singular, BERRY_2017__HalfCentury, Berry2020_JMP}.  Examples include an optical vortex beam, being a typical case of phase singularity discovered to carry an orbital angular momentum of light~\cite{ALLEN_2003__Optical, desyatnikov2005optical}, and vortices of polarization states (polarization singularities) in cylindrically polarized laser beams~\cite{zhan2009cylindrical} enabling tighter focusing~\cite{tovar1998production, youngworth2000focusing, zhan2002focus, dorn2003sharper}. Moreover, singularities of wave optics can be employed to facilitate  optical communications~\cite{cheng2009propagation, milione2015using, ndagano2017creation} and enhance light-matter interaction~\cite{zhan2004trapping, meier2007material, marago2013optical}.

Over the past decades, studies on wave optics have been extended to the microscale and even nanoscale realm, yielding the new field of {\em metaphotonics} that enables unprecedented manipulations of light-matter interactions.
Metaphotonics is a novel direction of research regarded as metamaterial-enabled photonics driven by strong optically-induced magnetic resonances of nonmagnetic structured elements~ \cite{koshelev2020dielectric}. Metaphotonics combines the concepts of metamaterials and nanophotonics, and it offers a unique platform for achieving unusual electromagnetic response on a scale much less than the wavelength, although several practical applications are strongly hampered by device fabrication complexity. As has been realized recently, optical singularities, especially polarization singularities, play essential roles in such structures, and they have triggered surging research enthusiasm.

In order to illustrate the current state of the field of metaphotonic-driven polarization singularities and predict its development in the future, here we review the recent discoveries concerning topological polarization singularities in nanophotonic structures and metasurfaces. More specifically, in Sec. 2 we discuss the polarization singularities associated with individual multipoles and their combinations manifested in far-field radiation. Section 3 covers the polarization singularities of the Bloch eigenmodes supported by metaphotonic structures in the far field.
%More specifically,
Sec. 3.1 reviews the nondegenerate polarization singularities and their link with bound states in the continuum (BICs), Sec. 3.2 surveys the degenerate cases and non-Hermitian effects, and then Sec. 3.3 discusses some works studying conversions from polarization singularities to phase singularities. Section 4 introduces both manipulation and observation of polarization singularities in the near-field of nanoscale structures. Finally, Sec. 5 concludes the paper.

\section{Polarization singularities and electromagnetic multipoles}

Currently, many different types of metamaterials and metasurfaces are being designed to modulate the far-field properties of radiation. In these structures, particles made of metals, high-index dielectrics or their hybrids are usually chosen to be the building blocks. Basically, to clarify the far-field  radiations of finite scattering bodies in terms of polarization singularities is crucial for the understanding of many related phenomena. For those finite structures, individual scattering particles in particular, Mie theory is well-suited to be employed to analyze the far-field behavior of the resonances supported. The fields of resonances can be expanded into the complete and orthogonal sets of vector spherical harmonics~\cite{jackson1962classical, bohren1983absorption, doicu2006light, powell2017interference, liu2018generalized}. Different multipolar moments (including electric and magnetic multipoles of various orders) have different far-field radiation patterns.

\begin{figure}[htpb]
\includegraphics[scale=.95]{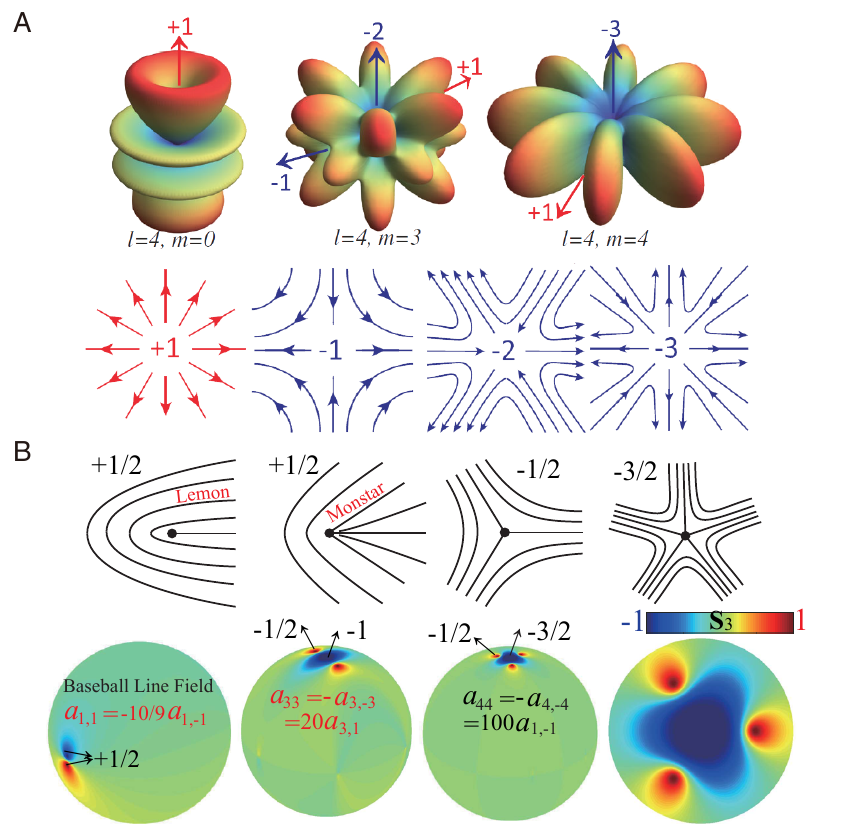}
\caption{(A) (Up) intensity null points in far-field radiation patterns of individual multipoles. (Down) examples of polarization singularities in the instantaneous vector fields of multipoles. (B) (Up) examples of polarization singularities in the  polarization fields of composite multipoles. (Down) existence of polarization singularities with half-integer Hopf indices  in their polarization fields. (A) is adapted from \cite{chen2019singularities} while (B) is adapted from \cite{chen2020line}.}
\label{fig:2}
\end{figure}

In the Mie theory, the subject of radiation directionality enhancement with proper design of the interferences among different multipolar radiation channels has attracted enormous attention~\cite{liu2018generalized, Kerker1983_JOSA, jahani_alldielectric_2016, KUZNETSOV_Science_optically_2016} (see also references therein). At the same time, there is actually another phenomenon worth detailed discussions. That is, there will be directions along which the radiation vanishes (non-radiative singularities, which are also special sorts of polarization singularities characterized by integer indexed or topological charges). Moreover, the radiation intensity will be greatly suppressed in the vicinity of those non-radiative directions. This phenomenon would lead to a possible effect: the more non-radiative singularities there are in the far field, the more suppressed the radiation would be, and thus the larger the quality factor (Q-factor) of the resonance. Such a effect serves as the underlying origin of high Q-factor supercavity modes in subwavelength dielectric resonators~\cite{rybin2017high}, the radiations of which can be represented by a combination of various multipoles. The hypothesis was firstly briefly mentioned in Ref.~\cite{bogdanov2019bound} and clarified systematically in Ref.~\cite{chen2019multipolar}. In these studies, Q-factor enhancement was observed when the dominant multipolar moment of the supermodes is converted from lower to higher order, accompanied by an increase in number of suppressed intensity points. Koshelev~\textit{et al.} in \cite{koshelev2020subwavelength} applied the Q-factor enhancement effect to boost the second-harmonic-generation efficiency of AlGaAs nanoparticles. In their work, with the dominant multipolar moment converted from a dipole to an octupole, the Q-factor of the resonance is enhanced by several orders.

The far electromagnetic fields of radiations are intrinsically transverse. As a result, the instantaneous far field of an individual linearly polarized multipole can be represented by a tangent real-vector field on a momentum sphere. In fact, the intensity null points in the far field are one kind of polarization singularities, stationary disclination singularities (vector polarization singularities or V-points) \cite{nye1983polarization}. In $2017$, Garcia-Etxarri noticed this important fact and conducted analytical calculations of the vector fields scattered by a high-index nanoparticle \cite{garcia2017optical}. The calculations showed that the two non-radiative poles in the far field of the dominant electric (or magnetic) dipole moment correspond to two stationary disclinations. A later paper by Chen \textit{et. al.} in \cite{chen2019singularities} gave a systematical analysis on the far-field V-point distributions for multipoles of various orders. It is found that there can be V-points both at the poles and away from the poles. The real-valued expression of the vector spherical harmonic $\mathbf{M}_{lm}$
\begin{equation}\label{eqn:realVecHarmonic}
  \begin{aligned}
    \mathbf{M}_{lm} = -\sin(m\phi) \pi_{lm}(\theta) \hat{\mathbf{e}}_\theta \\
    -\cos(m\phi) \tau_{lm}(\theta) \hat{\mathbf{e}}_\phi,
  \end{aligned}
\end{equation}
where $\tau_{lm}(\theta) = \frac{d}{d\theta}P_l^m(\cos\theta), \pi_{lm}(\theta) = \frac{m}{\sin\theta} P_l^m(\cos\theta)$, and $P_l^m(\cos\theta)$ denotes associated Legendre
polynomials, is chosen to characterize the electromagnetic field of a magnetic multipole. For such a representation, there is no phase difference between the vectors and the radiated light is linearly polarized everywhere. Defining the Poinca\'re index (or equivalently the topological charge) $\mathbf{Ind}_p$ as the winding angle of the field vectors around the singular point divided by $2\pi$, it is found the indices of V-points at the poles are
\begin{equation}\label{eqn:IndicesPole}
  \mathbf{Ind}_p = 1 - \abs{m},
\end{equation}
while the indices and positions of other V-points would be
\begin{equation}\label{eqn:IndicesNonPole}
  \begin{aligned}
    \mathbf{Ind}_p = -1, \cos(m\phi) = \pi_{lm}(\theta) = 0, \\
    \mathbf{Ind}_p = +1, \sin(m\phi) = \tau_{lm}(\theta) = 0.
  \end{aligned}
\end{equation}
From Eq. (\ref{eqn:IndicesNonPole}) it is clear that there would be $2\abs{m}(l-\abs{m})$ V-points with index of -1 and $2\abs{m}(l-\abs{m} + 1)$ V-points with index of +1. Summing all the indices on the momentum sphere, the total index will always be +2, which agrees with the Poincar\'e-Hopf theorem \cite{milnor1997topology, richeson2009euler}. %(more precisely the hairy ball theorem for this case): the total Poincar\'e index of a closed manifold (the momentum sphere here) equals to the Euler characteristic of the manifold (+2 here for a sphere).
On the other hand, the result of an electric multipole $\mathbf{N}_{lm}$ can be similarly derived from the expression of $\mathbf{M}_{lm}$ above considering the duality of Maxwell equations and the tangent field premise, transforming the basis: $\hat{\mathbf{e}}_\phi \rightarrow -\hat{\mathbf{e}}_\theta, \hat{\mathbf{e}}_\theta \rightarrow \hat{\mathbf{e}}_\phi$. Such a transformation will neither change the position nor the index of the singularity and index distributions. Consequently, both electric and magnetic multipoles with the same quantum numbers $(l,m)$ have the same index distributions. Fig. \ref{fig:2}A adapted from \cite{chen2019singularities} illustrates examples of the studied multipolar far-field patterns and the corresponding vector field patterns around the V-points. The authors also discussed the V-points in the vector field of combined multipoles which are kept to be linearly polarized, finding that a combined multipole such as (the Kerker dipole or a quadrupole) has only one singularity of index $+2$.

The aforementioned topic went further in \citen{chen2020line}, extended from real vectors to complex vectors (also linear polarizations to more general elliptic polarizations) that are characterized by line fields rather than vector fields. The same Poincar\'e-Hopf theorem was therefore applied to these line-field singularities \cite{delmarcelle1996visualization, hopf2003differential}. Defined as the winding angle of polarization major (or minor) axes divided by $2\pi$, the topological indices (called Hopf indices, $\mathbf{Ind}_h$) of singularities in a line field can be half integers. The first row of Fig. \ref{fig:2}B gives some examples of line singularities with different Hopf indices. Importantly, a line singularity can not only be a V-point which we have introduced above, but also a circularly polarized point (C-point) where the field oscillates in a circular orbit. A C-point and a V-point are indistinguishable in a line field, but in the actual complex-vector field they are totally different. For a V-point, the ill-defined direction of the polarization state is accompanied by a zero intensity. In contrast, for a C-point, though the azimuthal angle of the polarization state is as well ill-defined, the intensity does not vanish. Moreover, C-points cannot be defined in an instantaneous real-vector field. Instead of V-points, C-points of indices of $\pm 1/2$ are generic singularities in the complex vector field, and V-points of integer indices can be broken into a series of C-points by further perturbations.

In order to study the far-field complex-vector field in an analytical way, complex-valued expressions of spherical harmonics $\mathbf{N}_{lm}$ (isolated electric multipoles) and $\mathbf{M}_{lm}$ (isolated magnetic multipoles) can be adopted as the set of basis:
\begin{equation}\label{eqn:complexVecHarmonic}
  \begin{aligned}
    & \mathbf{N}_{lm} = [\tau_{lm}(\theta) \hat{\mathbf{e}}_\theta + i \pi_{lm}(\theta) \hat{\mathbf{e}}_\phi]
    \frac{[krz_l(kr)]'}{kr} \exp(im\phi), \\
    & \mathbf{M}_{lm} = [i \pi_{lm}(\theta) \hat{\mathbf{e}}_\theta - \tau_{lm}(\theta) \hat{\mathbf{e}}_\phi]
    z_l(kr) \exp(im\phi).
  \end{aligned}
\end{equation}
One would find that behaviors of the far field of fundamental multipoles being studied here are pretty different from the previously studied real-valued ones (actually they are complex combinations of the `new' basis here). %The `new' basis chosen here is a bit more special than the previous one that the complex harmonics don't have time-reversal symmetry in general.
Looking into Eq. (\ref{eqn:complexVecHarmonic}), the conclusion can be drawn that, for arbitrary order of multipolar moment there can only be two isolated polarization singularities with Hopf index $\mathbf{Ind}_h$ equals to $+1$, which might either be C-points or V-points, at the two poles of the far field. To obtain V-points or C-points with other Hopf indices like $\pm1/2$, combinations of the complex multipolar basis are necessary, for which some specific examples are shown in the second row of Fig. \ref{fig:2}B. The sum of Hopf indices over the momentum sphere are still secured by Poincar\'e-Hopf theorem to be $+2$ for all cases.

Those studies provide insights into the relation between the properties of local resonances supported by finite scattering bodies and the far-field landscapes of polarization singularities. Since particles are building blocks of metamaterials, the multipolar perspective based on polarization singularities serves as a new viewpoint to clarify complicated phenomena in metaphotonics, as will be further discussed in the next sections.

\section{Two-dimensional periodic photonic structures}

As introduced in the previous section, polarization singularities have been observed in the far field of finite scattering bodies that can serve as building blocks of extended metaphotonic structures. Those questions thus naturally arise: will there also be optical singularities in the far field of infinite metaphotonic structures? Will those singularities lead to interesting phenomena or applications? The answers are yes. In this section, we focus on two dimensional (2D) periodic structures composed of nano-particles or holes as unit cells. Basing on the radiative Bloch modes, 2D periodic structures can be employed to flexibly modulate light propagation and the polarization state.

\subsection{Radiative Bloch modes without degeneracies of photonic bands }

Two-dimensional periodic photonics structures, \textit{e.g.} photonic crystal slabs (2D PhC slabs), plasmonic crystals, and periodic metasurfaces, are the structures composed of identical units forming a two-dimensional lattice. Due to periodicity, both localized and propagating modes in the structure will turn into Bloch modes with well-defined in-plane Bloch wave vectors $k_\parallel$. The Bloch modes will form photonic bands, and the part of the bands above the light cone is radiative, producing radiations to the free space. In last section, we have shown that polarization singularities are generic in the far field of finite scattering bodies that are building blocks of metaphotonic structures. Since the optical response of a 2D periodic structure can be effectively viewed as the collective behavior of its building blocks (after considering the coupling among different unit cells), it is expected that there will be polarization singularities for Bloch modes in the far field too.

%Compared to gradient metasurfaces which can easily induce both phase and polarization singularities by inhomogeneity \cite{pancharatnam1956generalized, berry1984quantal, berry1987adiabatic, bomzon2002space, yu2011light, karimi2014generating, yang2014dielectric, chen2015generation, pu2015catenary, hakobyan2016tailoring, yue2016vector, devlin2017spin, tang2019high}, 2D periodic structures are relatively more homogeneous and do not exhibit strong spatial anisotropy. As a consequence, 2D periodic structures are not as flexible to tune the polarization or phase distributions of propagating light. Nevertheless, it doesn't draw the conclusion that no polarization singularities would appear in the far field of those structures.

Early observations of polarization singularities in the far fields of 2D periodic structures are associated with bound states in the continuum. In 2014, Zhen Bo~\textit{et al.} \cite{zhen2014topological} firstly reported the existence of vector polarization singularities in the radiation far field of PhC slabs. It is worth mentioning that the definition of the far field in this work is a little different from those in conventional studies, where the frequency is variant rather than fixed. To be more specific, the far field here is defined on one specific photonic band, which means the frequency is varying with changing in-plane wave vectors $k_\parallel = (k_x, k_y)$. Polarization state of each Bloch resonant mode on the studied band can be mapped onto the defined far field (or the momentum space), constructing a far-field polarization field. Vector polarization singularities (V-points as mentioned previously) may appear in this polarization field and radiation intensity at those singularities would vanish. The corresponding states will therefore be non-radiative although they fall into the radiative continuum (the light cone), \textit{i. e.} they are bound states in the continuum. The band taken as an example was the second TM-like band of a PhC slab. Five V-points can be found in the simulated polarization field, and the $Q$-factor map directly proves the corresponding BICs, of which the Q-factors tend toward infinity. Analyzing the relationship between the symmetry and the polarization field, the authors concluded that the index of the central pinned V-point is determined by the symmetry of the system: the higher the fold of rotational symmetry is, the larger the index may be. In addition, they also showed the movement, annihilation and emergence of the V-points (also BICs) in PhC slabs with other symmetries by numerical calculations.

\begin{figure}[htpb]
\includegraphics[scale=.95]{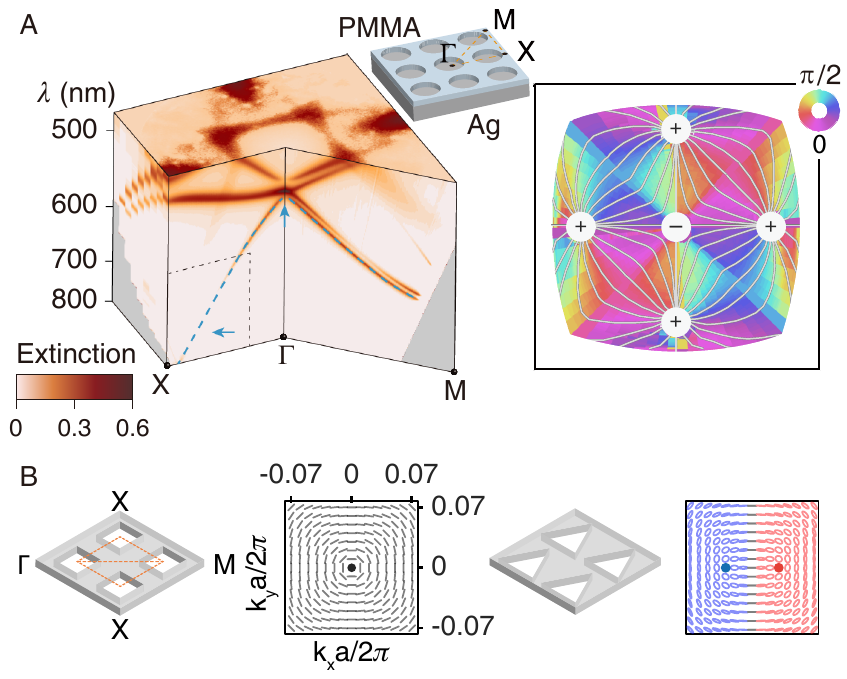}
\caption{(A) Direct observation of polarization singularities in the far field of a square-lattice plasmonic crystal in the form of absorption spectra. Left: the measured band structure of the plasmonic crystal. Right: the reconstructed far-field polarization field on band 1 of the plasmonic crystal, showing existence of 5 vector polarization singularities. (B) An example of spawn of circularly polarized states (C-points) from a vector polarization singularity (V-point) in the momentum-space polarization field of a photonic crystal slab. When the holes are transformed from squares to isosceles triangles, the V-point correspondingly separated into two C-points. (A) is adapted from \citen{zhang2018observation} and (B) is adapted from \citen{liu2019circularly}.}
\label{fig:3}
\end{figure}

Afterwards in 2018, Zhang \textit{et al.} \cite{zhang2018observation} made the first direct observation on the far-field V-points on photonic bands in a $C_{4v}$ symmetric plasmonic crystal (see Fig. \ref{fig:3}A). Applying a Fourier-transformation-based spectroscopy, they were able to measure the angle-resolved extinction spectra, which showed the band structure, as plotted in the first panel. Analyzing the measured extinction which is reciprocal to radiation, they mapped out Q-factors of resonances on SPP band 1, and reconstructed the polarization field considering not only the polarization azimuths (see second panel of Fig. \ref{fig:3}A) but also the polarization ellipticity. Five V-points (also BICs) were observed. They also exhibited the polarization-resolved band surfaces obtained by summing the iso-frequency extinction maps, which directly visualized the winding polarization states in the vicinity of the V-points as dark patterns spinning with the analyzer.

The same year, another work also proved the existence of vector polarization singularities on the photonic band of a grating \cite{doeleman2018experimental}. They found that the reflected polarization states near the BIC on the studied photonic band also appeared to be winding. As a supplement to the papers mentioned \cite{zhen2014topological, zhang2018observation}, a note from C. W. Hsu \textit{et al.} \cite{hsu2017polarization} gave the theoretical proof that the polarization states of radiation from 2D periodic structures are generally elliptical. However, with an in-plane $C_2$ symmetry, the ellipticity of the polarization states would be close to 0. In 2019, Liu \textit{et al.} \cite{liu2019circularly} introduced symmetry breaking into Si$_3$N$_4$ PhC slabs, and found that the previously studied V-points would be perturbed into more elementary polarization singularities, the C-points. With the square holes in the studied slabs gradually transformed to isosceles trapezia and then isosceles triangles, the in-plane $C_2$ symmetry of the system is further broken, resulting in the spawning of C-points and the rise of ellipticity of the polarization states in the far field (see Fig. \ref{fig:3}B). The authors also explored the possibilities to modulate the momentum-space polarization field with different types of symmetry breaking.

About the origin of the polarization singularities in the far field defined on photonic bands, the multipolar expansion perspective introduced in the previous section \cite{chen2019singularities, sadrieva2019multipolar, chen2020line} can be applied, though the modes may be extended states rather than localized resonances. Bloch modes in 2D periodic structures can be expanded into periodically distributed multipole moments, and far-field singularities of the structures are revealed to stem from the ones in the far field of the combinations of expanded multipoles. As examples, Chen \textit{et. al.} \cite{chen2019singularities, chen2020line} gave the far fields of the multipoles expanded from the $\Gamma$-point BICs on the second TE-like band and the second TM-like band in a $C_{2v}$-symmetric PhC slab, and also mapped the ellipticity distributions of the far field obtained by expanding specific modes on the second TE-like band of a $C_{2v}$-symmetry-broken PhC slab. Both the examples given show obvious correspondence between the polarization singularities in the expanded radiation field and the ones in the far field defined on photonic bands. The intensity null points (V-points) in the expanded radiation patterns forced by symmetry correspond to the symmetry-protected $\Gamma$-point BICs, and the mobile off-$\Gamma$ BICs originate from the null-points caused by destructive interference. Meanwhile, the C-points originating from complex combinations of multipoles correspond to the circularly polarized modes near the $\Gamma$ point as a result of $C_2$ symmetry breaking.

\begin{figure*}[t!]
\includegraphics[scale=.95]{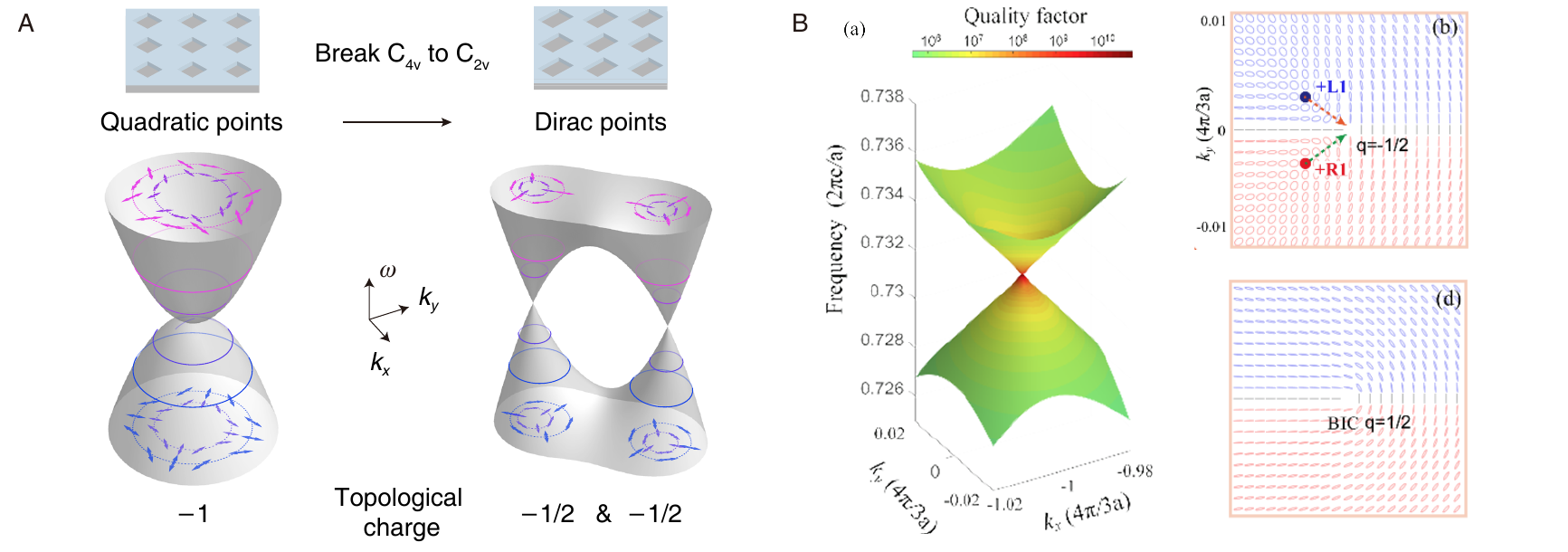}
\caption{(A) Schematic views of polarization singularities at band degeneracies. A quadratic degeneracy will have a radiative integer-index V-point, while a linear degeneracy (Dirac point) will have a radiative half-integer-index V-point. A quadratic degeneracy is protected by rotational symmetry higher than 2 folds, and by breaking the symmetry of a plasmonic crystal from $C_{4v}$ to $C_{2v}$ (see insets), it will split into two linear degeneracies, and the corresponding V-point would split as well. (B) An example of naturally existing C-points near a Dirac point at the corner of a hexagonal-lattice photonic crystal slab. Their mergence with the radiative V-point at the Dirac point would result in a non-radiative V-point, i. e. a BIC. Left: the quality factor map of the Dirac degeneracy. Right: the polarization map on the upper band of the degeneracy, before and after the mergence. (A) is adapted from \citen{chen2019observing} and (B) is adapted from \citen{ye2020singular}.}
\label{fig:4}
\end{figure*}

The paper~\citen{liu2019circularly} is, to the best of our knowledge, the very first study in which the generic polarization singularities (that are C-points rather than non-generic V-points) into the far field of metaphotonic structures. Following this work, a lot of studies concerning polarization singularities in metaphotonics have been conducted. Ye \textit{et al.} reported the existence of C-points near the boundary of the first Brillouin zone of a hexagonal PhC slab which may be related to diffraction \cite{ye2020singular}. Yin \textit{et al.} discovered that, by re-merging two C-points separated from a $\Gamma$-point V-point (BIC) in a PhC slab with a substrate, a guided resonance which radiates to only one direction can be realized~\cite{yin2020observation}. Guo~\textit{et al.} showed that by lifting the Dirac degeneracy at the Brillouin zone edge of a honeycomb PhC slab, C-points and meron spin textures would be generated on the valleys \cite{guo2020meron}. Yoda \textit{et al.} gave further results on manipulations of polarization singularities in the far field of 2D periodic structures, in which generation of six C-points from one V-point was shown possible with the symmetry of system broken from $C_{6v}$ to $C_{3v}$ \cite{yoda2020generation}. The so-defined far-field polarization fields on photonic bands may contain information of band topologies and topological properties of the system~\cite{fosel2017lines}, which brings a new vision to the research of metaphotonics and topological photonics. Besides, due to the high-Q nature of the far-field V-points, 2D periodic structures may potentially be employed to produce vectorial laser beams~\cite{kodigala2017lasing, ha2018directional, huang2020ultrafast, wang2020doubly}.

\subsection{Radiative Bloch modes with degeneracies of photonic bands}

We introduced a far-field polarization field defined on one photonic band of a 2D periodic structure in last section. One would find that, most discussions on such momentum-space polarization field concentrated on singlet states. On account of the common existence of degeneracies in 2D periodic structures and their unique properties, the discussion should be extended to degenerate states. A paper from Chen \textit{et al.} \cite{chen2019observing} discussed about this topic. The authors argued that the previous definition of polarization fields on photonic bands shall still be applicable with the existence of degeneracies. Obviously, polarization states of all the Bloch singlet modes on the photonic bands are still well-defined under this definition. Whereas, a degeneracy is a superposition of multiple modes, so its polarization state is automatically undefined. Consequently, as the touching points of bands, the degeneracies can be naturally treated as singular points of the polarization field.

The authors also proposed that the topological charges of the degenerate polarization singularities may be associated with the symmetries protecting them, hence can reflect the symmetry-protected local topological properties of the degeneracies. In Fig. \ref{fig:4}A, the authors plotted the two general classes of degenerate states in 2D periodic structures with corresponding schematic polarization configurations: quadratic degeneracies and linear degeneracies (diabolic or Dirac points). Quadratic degeneracies may be protected by rotational symmetries of order higher than 2 at high-symmetry points of Brillouin zone. Meanwhile, Dirac degeneracies may be protected by a product of $C_2$ and time-reversal ($T$) symmetry. For quadratic degeneracies which are trivial in topology, a closed loop around the degenerate point on either degenerate band will introduce trivial geometric phase. Consequently, tracing around such a loop, the polarization state can return to itself with no effective phase shift, \textit{i. e.} the polarization singularity shall be an integer-charged point.

In contrast, the loop around a Dirac point is accompanied by a $\pi$ Berry phase, making the charge of the polarization singularity to be a half-integer. This can be viewed as a Bloch analogue of the Hamilton's diabolo in biaxial crystals \cite{hamilton1837third, born1959principles, berry2007conical, leykam2016conical}. It is worth emphasizing that, the two classes of degeneracies can be connected by symmetry breaking: a quadratic degeneracy will split into two Dirac points when the symmetry of system change from $C_{nv}$ to $C_{2v}$ (with $T$ symmetry not broken) \cite{chong2008effective}. The authors utilized this connection to show the relationship between the polarization singularities carried by quadratic degeneracies and Dirac degeneracies (see Fig. \ref{fig:4}A). Along with the split of Dirac degeneracies from quadratic degeneracies, the corresponding integer vector polarization singularity would split into a pair of half-integer singularities. The authors verified the proposed theory by simulations of a plasmonic crystal. They also observed the singularities with the same method as in \cite{zhang2018observation}, and discussed the case of accidental degeneracies. Later, in work from Ye \textit{et al.} \cite{ye2020singular}, the authors showed cases that vector polarization singularities (or V-points) also appear at the Dirac points on the edges of the first Brillouin zone in a honeycomb-lattice PhC slab. The complex interference between radiation channels beyond the diffraction limit can also lead to natural existence of circularly polarized states (or C-points). Their mergence with the radiative V-points at the Dirac points will change the topological indices of the V-points and turn those V-points to non-radiative ones (BICs). Figure \ref{fig:4}(b) shows the simulated quality factor mapped onto the degeneracy when the mergence happened, and the corresponding polarization maps on its upper band before and at the time of mergence are also shown.

\begin{figure}[h!]
\includegraphics[scale=.95]{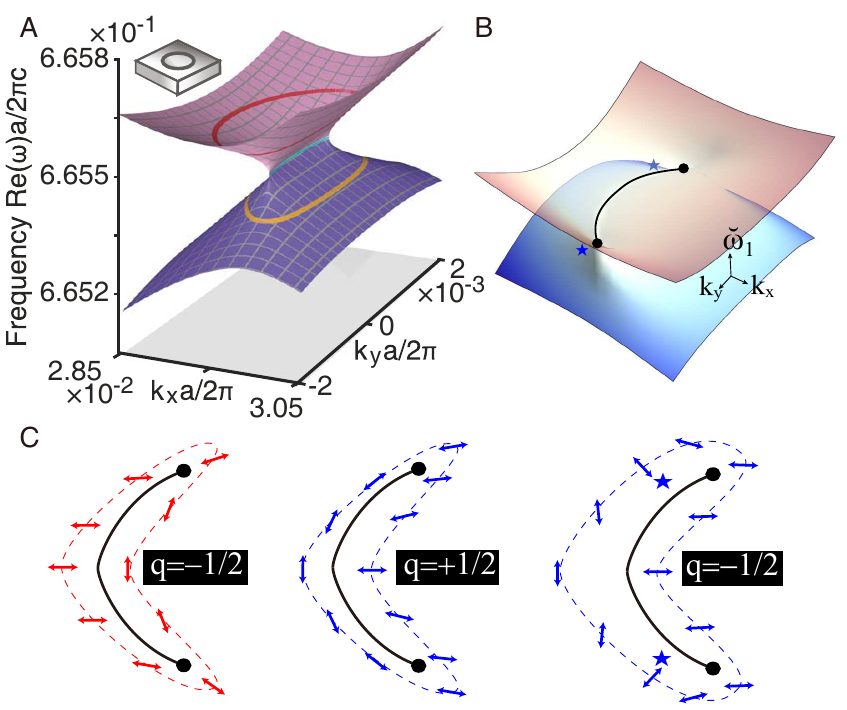}
\caption{(A) An example of a Dirac degeneracy turning into a bulk Fermi arc due to non-Hermiticity. The system is a rhombic-lattice photonic crystal slab. (B) The existence of circularly polarized states near the two ends of the bulk Fermi arc (the two exceptional points) on the lower band. (C) Polarization state distributions on different iso-frequency contours around the bulk Fermi arc. For the upper band, the winding number of polarization states equals to $-1/2$ which is expected. Meanwhile, for the lower band, if the two C-points are not enclosed by the chosen iso-frequency contour, the winding number can unexpectedly be $+1/2$. (A) is adapted from \citen{zhou2018observation}, while (B) and (C) are adapted from \citen{chen2020evolution}.}
\label{fig:5}
\end{figure}

A non-negligible fact about the degeneracies is that, the corresponding irreducible representations (irreps) of some degeneracies may match the irreps of free-space plane waves. As a result, these degeneracies can be radiative and are able to be excited by arbitrary incident polarization, although they are also V-points like BICs. On the contrary to the non-radiative BICs, the resonance-lifetime will be pretty low for the radiative degeneracies. The strong radiation loss will induce intense non-Hermiticity to the degeneracies, leading to non-Hermitian phenomena such as exceptional rings spawning from accidental Dirac points \cite{zhen2015spawning}. In the more general case of symmetry protected Dirac points, the non-Hermiticity would make the Dirac degeneracies become bulk Fermi arcs (arcs where the real part of eigen-frequencies degenerate) and lead to exceptional points (where two or more resonances coalesce in both eigen-frequencies and eigenfunctions), as reported in the seminal paper by Zhou \textit{et al.} \cite{zhou2018observation}. By theoretical analysis, they found that in a $C_{2v}$-symmetric rhombic-lattice PhC slab, there will be two exceptional points as ends of the bulk Fermi arc on specific photonic bands (see Fig. \ref{fig:5}A). The bulk Fermi arc originates from eigenvalues of a complex effective Hamiltonian perturbed by radiative losses from the hamiltonian of a Dirac point. The two exceptional points are the points where the complex Hamiltonian come to have degenerate eigen-frequencies and eigenmodes. In the polarization field defined on photonic bands, the signature of a bulk Fermi arc was found to be a half-integer topological charge. This feature seems well-reasoned since bulk Fermi arcs come from Dirac points which correspond to half-integer singularities as discussed \cite{chen2019observing}. The authors fabricated the designed sample and confirmed experimentally the existence of bulk Fermi arcs and exceptional points through an open-ended iso-frequency contour and the measured half-integer ($-1/2$) polarization winding number on iso-frequency contours around the open-ended Fermi arc.

In both Ref.~\citen{zhou2018observation} and Ref.~\citen{chen2019observing}, the researchers only showed sketchy polarization maps around the non-hermitian degeneracies. It makes people curious about the detailed polarization behaviors in the vicinity of a bulk Fermi arc. In the work by Chen~\textit{et al.}~\cite{chen2020evolution}, the authors conducted a careful theoretical study on the polarization field around those non-hermitian degeneracies. They applied the previously studied \cite{zhou2018observation} rhombic-lattice PhC slab to map out the polarization field around the Fermi arc, discovering that there were actually two C-points appearing in the polarization field defined on the lower degenerate band (see Fig.~\ref{fig:5}B). The detailed map as Fig. \ref{fig:5}C shows an interesting fact that, the polarization charge of a Fermi arc may deviate from the prediction that it shall be half the integer charge of the corresponding quadratic degeneracy~\cite{chen2019observing}. The original quadratic degeneracy here have a polarization singularity of $-1$, and are expected to be splitted into Fermi arcs with $\mathbf{Ind}_h = -1/2$ polarization singularities. The first panel of Fig. \ref{fig:5}C shows that the polarization states on the upper band are well predicted. However, for the polarization states on the lower band which are marked by blue arrows in the other two panels of Fig. \ref{fig:5}C, the winding number on an iso-frequency contour enclosing the arc can be either $-1/2$ or $+1/2$. It is only by closing the two C-points that the Hopf index of the Fermi arc can agree with the prediction. Further breaking the mirror symmetry of the system, they found that there would even be cases that the polarization-azimuth winding angle on an iso-frequency contour enclosing the Fermi arc is not equal to the invariant Berry phase. Even so, the authors observed that the global topological index of the polarization fields on the degenerate bands would still be conserved, mediated by the C-points. They pointed out that the polarization charge conservation is consistent with the Berry phase invariance. Analytic Hamiltonian models were applied in order to verify the theory the authors proposed explaining the observed phenomena.

The studies on momentum-space polarization fields near non-Hermitian degeneracies are important steps towards the application of singular optics in studying topological and non-Hermitian phenomena inside metaphotonic structures. There is also a research showing the existence of true diabolic degeneracies in non-Hermitian systems with specific symmetries \cite{xue2020non}, indicating that there are more interesting phenomena to be revealed associated with tuned symmetries and non-Hermiticity. Blending the concepts from various fields including singular optics, topological photonics, and non-Hermitian physics with metaphotonics, more and more exotic phenomena and refreshing applications are expected to be inspired.

\subsection{Conversion from polarization singularities to phase singularities}

\begin{figure*}[t!]
\includegraphics[scale=.95]{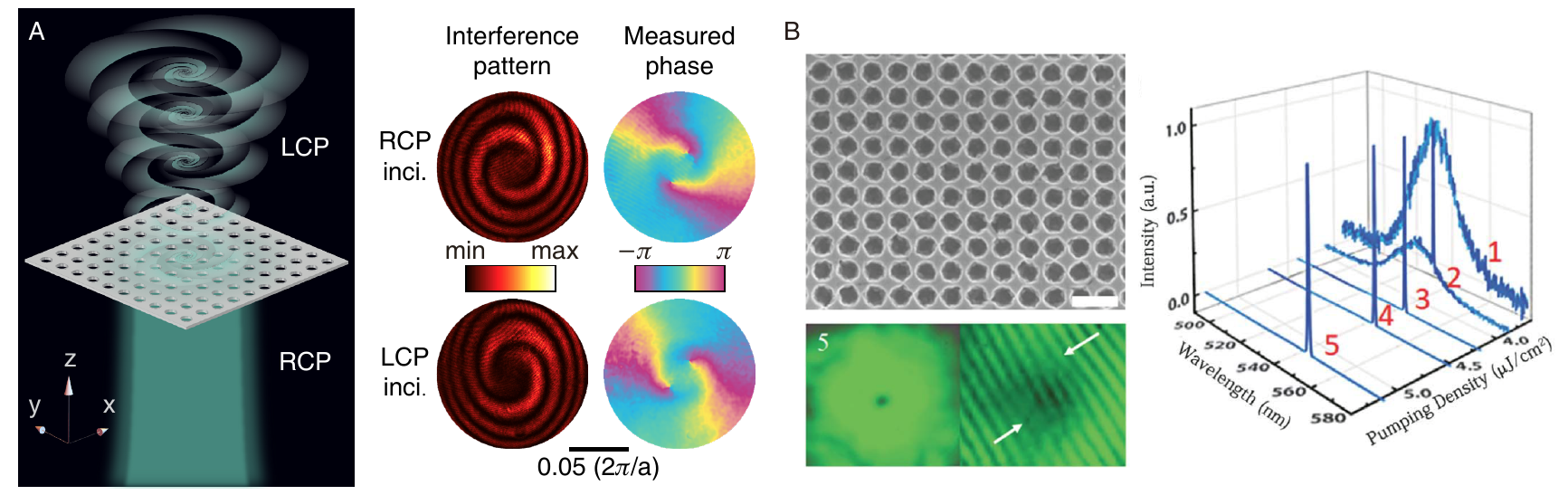}
\caption{(A) Left: the schematic view of generating an optical vortex beam with a photonic crystal slab. When a slightly convergent beam with a certain circular polarization and the chosen working wavelength is shined on the photonic crystal slab, the resonance-converted cross-polarized beam radiating from the slab would gain a spiral phase front. Right: measured interference patterns and extracted spiral phase distributions in the validating experiments. (B) A perovskite-based photonic-crystal-slab sample which can actively generate a vortex laser. With the increasing pumping power density, the sample clearly exhibits a lasing action as plotted in the right panel. The torus-shaped beam profile and the interference pattern on the left show the vortex behavior of the generated laser beam. (A) is adapted from \cite{wang2020generating}, and (B) is adapted from \cite{huang2020ultrafast}.}
\label{fig:6}
\end{figure*}

We notice that the polarization field introduced above is actually a field indicating the polarization responses of resonant modes of a 2D periodic structure in the momentum space. Readers might be reminded of the mechanism of Pancharatnam-Berry (PB) phase based optical vortex generators: spatial winding polarization responses of resonators are utilized to induce helical geometric phase shifts to the incident light \cite{pancharatnam1956generalized, berry1984quantal, berry1987adiabatic, bhandari1997polarization, bomzon2002space, yu2011light, karimi2014generating, yang2014dielectric, chen2015generation, pu2015catenary, hakobyan2016tailoring, yue2016vector, devlin2017spin, tang2019high, bliokh2019geometric}. Will momentum-space winding polarization responses be capable of generating phase singularities in the far field as well? The paper by Wang \textit{et al.} \cite{wang2020generating, notomi2020topology} gave the definite answer. The authors proposed a new way to generate optical vortices with 2D periodic structures, implementing the spin-orbit interaction (SOI) in momentum space, as shown in the upper panel of Fig. \ref{fig:6}A. Let's consider a parabolic photonic band in a 2D periodic structure. If there is a V-point (BIC) at its center, the far-field polarization states of the resonant modes, of which the frequencies are close to the BIC, will be winding on the iso-frequency contours. The winding number will be equal to the topological charge (or Hopf index) $q$ of the central V-point. Now, a circularly-polarized Gaussian beam of a specific frequency, whose momentum-space distribution covers the corresponding iso-frequency contour, is shone onto the 2D periodic structure. Part of the beam will be converted into the other handedness due to the Bloch resonances. The unconverted part of the beam is then dropped by applying a cross-polarized analyzer, keeping only the converted beam. The polarization states of different $k$-components in the transmitted beam will have experienced different changes according to different polarization responses of the resonances with different $k_\parallel$, which can be deduced using Jones calculus with a helical polarization basis:
\begin{equation*}
  \begin{aligned}
    \left| E_{\mathrm{out}} \right> =
    &
    \frac{1}{2}
    [t_x(\mathbf{k_{\parallel}}) + t_y(\mathbf{k_{\parallel}})]
    \left| E_{\mathrm{in}} \right>
    + \\
    &
    \frac{1}{2}
    [t_x(\mathbf{k_{\parallel}}) - t_y(\mathbf{k_{\parallel}})]e^{-2i\theta(\mathbf{k_{\parallel}})}
    \braket{E_{\mathrm{in}}|R}
    \ket{L}
    + \\
    &
    \frac{1}{2}
    [t_x(\mathbf{k_{\parallel}}) - t_y(\mathbf{k_{\parallel}})]e^{2i\theta(\mathbf{k_{\parallel}})}
    \braket{E_{\mathrm{in}}|L}
    \ket{R}.
  \end{aligned}
\end{equation*}
Here $t_x, t_y$ are respectively the transmittance coefficients of the $\mathbf{k_{\parallel}}$ Bloch mode with the polarization parallel and perpendicular to the efficient fast axis; $\theta$ is the azimuthal angle of major axis of the polarization state to the $x$-axis; $\ket{E_{\mathrm{in}}}, \ket{E_{\mathrm{out}}}$ are the Jones vectors of the incident and transmitted light; while $\ket{L}, \ket{R}$ denote the left- and right-handed circularly polarized (LCP \& RCP) unit vectors $(0,1)^\mathrm{T}$ \& $(1,0)^\mathrm{T}$. From the above formula, it is clear that the output beam after the analyzer will be modified both in intensity and phase -- the intensity distribution will follow the shape of the iso-frequency contour, while a phase vortex with topological index $l = \pm 2\  \mathbf{Ind}_h$ will be introduced. The researchers fabricated a freestanding Si$_3$N$_4$ PhC slab, and applied the resonances at 532 nm near a BIC to verify their proposed theory. They measured intensity profiles, interference patterns, and phase distributions of the generated beams with circularly polarized incident light, confirming the existence of phase vortices, shown as Fig. \ref{fig:6}A.

The proposed principle may look identical to PB phase gradient metasurfaces. However, the base of the winding polarization responses applied in this theory are the momentum-space V-points rather than real-space designed resonators. Such polarization singularities, or BICs, exist naturally in 2D periodic non-gradient structures and were discovered in different structures made of various kinds of materials. The only conditions required are the symmetry and the periodicity. Thus, such a new principle of optical vortex generating would render more flexibilities. And as mentioned, the index $\mathbf{Ind}_h$ of V-points can be easily controlled by the rotational symmetry of the system, so one can purposefully design the structure to obtain desired order $l$ of optical vortex. Another important feature is that, such a principle removes the requirement of spatial inhomogeneity to induce SOI. As a result, no more spatial alignment of the structure is required according to the beam center. It is worth mentioning that a similar theory has also been proposed by Liu \textit{et al.} \cite{liu2020universal} in honeycomb and Lieb photonic lattices, linking the optical pseudo-spins to far-field optical vortices.

Such a mechanism not only applies to passive modulations of light, but also is valid in active lasing actions. The paper by Huang \textit{et al.} \cite{huang2020ultrafast} realized vortex microlasers within a square-lattice PhC slab made of perovskite (MAPbBr$_3$). A perovskite film patterned with circular holes was fabricated and sandwiched by a substrate and a cladding layer of which the refractive index $\approx 1.5$ (see Fig. \ref{fig:6}B). The $\Gamma$-point vector singularity on the second band TM-like band of the sample leads to a BIC within the gain spectral range of MAPbBr$_3$. The fabricated sample was then pumped by a blue Ti:sapphire laser at room temperature. With low pumping density, a broad spontaneous emission peak centered at 520 nm appeared. Enhancing the pumping power, a sharp peak at 552 nm corresponding to the BIC appeared, and this peak quickly dominated the spectral signal with a further increase of the power. This phenomenon clearly marked the lasing action. The profiles and self-interference patterns of the laser beam were measured at different pumping powers, clearly exhibiting the existence of polarization and phase vortices (see Fig. \ref{fig:6}B). To Confirm the vortex nature of the laser beams, the authors made further efforts to experimentally prove the ability of all-optical control of the vortex lasing action. The profile of the pumping beam was modulated to be an ellipse or two overlapped circles, which would lead to a spatial redistribution of the gain coefficient. Consequently the symmetry of the PhC slab would be broken, and thus the output would turn out to be two lobe-shaped linear polarized beam, different from the original donut-shaped vortex beam. The time delay of such a switching was measured to be approximately $1.5$ ps.

Applying fundamental concepts of singular optics to the far-field polarization and phase behaviors of 2D periodic structures, important optical phenomena like BICs have been interpreted from a more general perspective. It is also inspiring researchers to explore exotic applications of 2D periodic structures which was previously considered more or less exhausted. Blending singular optics with periodic structures has certainly cross fertilized both disciplines.

\section{Near-fields generated by metaphotonic structures}

Beyond the far-field behavior of light, sophisticated metaphotonic structures can precisely modulate optical near fields. Polarization singularities are also attaining attention in those nano-scale near fields. These nanoscopic fine structures in the light field of a metaphotonic structure may be used to control the flow of light at nano-scale.

\begin{figure}[htpb]
\includegraphics[scale=1]{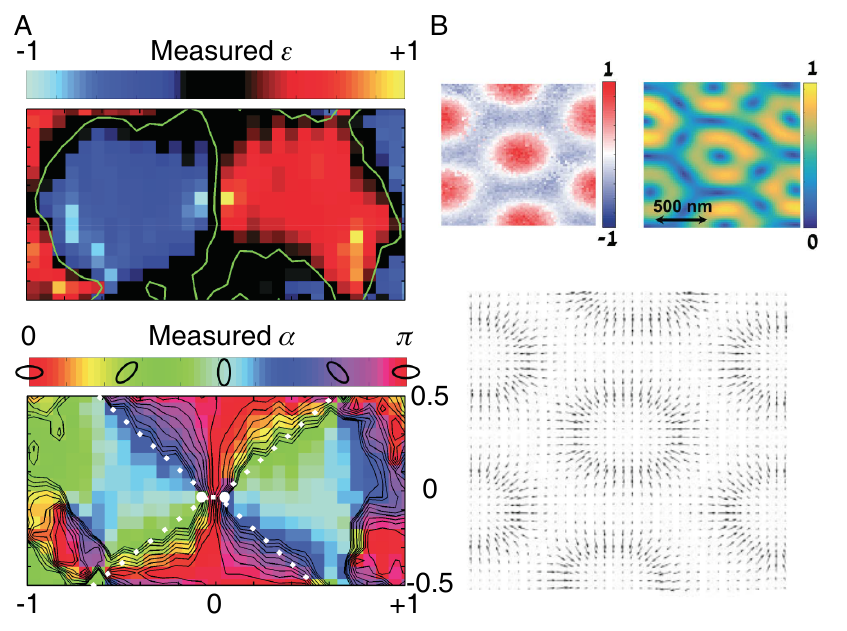}
\caption{(A) Observation of polarization singularities such as C-points and disclinations in the near field of a photonic crystal waveguide using a NSOM. The measured azimuthal angle map and ellipticity map both show the existence of polarization singularities. (B) An example of near-field polarization singularity manipulation. A lattice of V-points, called an `optical skyrmion lattice', was generated using a hexagonal-shaped plasmonic grating. (A) is adapted from \citen{burresi2009observation}, (B) from \citen{tsesses2018optical}.}
\label{fig:7}
\end{figure}

For example, C-points as the generic polarization singularity can be found among the fields inside a photonic crystal waveguide \cite{burresi2009observation, le2015nanophotonic, young2015polarization, lang2015stability, sotto2019spin} or around metallic holes \cite{de2014optical, de2015creating}. The radiation of quantum emitters inside the metaphotonic structures can be tuned by modifying their position according to those singularities \cite{le2015nanophotonic, young2015polarization}. On the other hand, Ostrovsky \textit{et al.} \cite{ostrovsky2018nanoscale} also demonstrated continuous nanoscale control over C-points in near field by changing the polarization state of a plane wave coupled to a plasmonic spiral slit. Another kind of polarization singularity we mentioned, vector polarization singularity (V-point), is also realized in the near-field of a hexagon-shaped plasmonic grating, named as `optical skyrmion' \cite{tsesses2018optical}. These works show the possibility of modulating light-matter interaction through fine polarization structures in metaphotonic devices.

Nevertheless, observing nano-scale singularities experimentally is intrinsically a challenging task. The basic problem is that, the abundant and various field behaviors at nano-scale in systems like plasmonic structures are in the near-field, which is evanescent and cannot be captured by imaging devices directly. To solve this problem, many methods were developed to offer an intermediate step converting part of the near field to far field aiming to observe the near field. Among them, the near-field microscopy technique based on near-field scanning optical microscopes (NSOMs) are the most optimal one since it will not alter the studied structure and can give access to as much information of the light field as possible. Aperture probes and scattering tips are used in those microscopes to extract the near-field information by converting the near fields to collectable radiations or scattering the near fields to the free space.

Since the invention of NSOMs \cite{synge1928xxxviii, pohl1984optical, lewis1984development, massey1985subwavelength, betzig1991breaking, trautman1994near}, they have been playing a significant role in measuring near-field intensity distributions and clarifing exciting phenomena in nano-scale optical fields~\cite{bozhevolnyi2002near, gersen2005direct, gersen2005real, lalanne2006interaction, zentgraf2008amplitude, esteban2008direct, rang2008optical, schnell2009controlling, deutsch2010visualizing, alonso2011real, chen2012optical, fei2012gate, le2014simultaneous}. To confer the capability of mapping the vectorial near field in NSOMs, interferometers and polarizers can be applied in the microscopes to achieve information about phase and polarizations~\cite{nesci2001quantitative, lee2007vector, kim2008polarization, olmon2008near, vignolini2009polarization, schnell2009controlling, schnell2010phase, alonso2012resolving}. Utilizing those techniques, researchers managed to observe different kinds of optical singularities in a variety of metaphotonic structures \cite{nesci2001optical, nesci2002measuring, dandliker2004measuring, fischer2008characterization, burresi2009observation, de2014optical, rotenberg2015tracking, spektor2015metafocusing, schnell2016real, ostrovsky2018nanoscale, tsesses2018optical, sun2019probing}. In 2009, polarization singularities were firstly experimentally measured by Burresi~\textit{et al.} \cite{burresi2009observation} in a photonic crystal waveguide using a polarization-sensitive NSOM. Not only C-points, but also linear polarization lines (L-lines) and disclinations were clearly observed and the distribution of polarization states in the photonic crystal waveguide were well mapped, as shown in Fig. \ref{fig:7}A. Further studies on nano-scale polarization singularities were then conducted with developed NSOM techniques \cite{de2014optical, rotenberg2015tracking}, including an observation on an ``optical skyrmion'' lattice \cite{tsesses2018optical} [shown as Fig. \ref{fig:7}B].

Studies about manipulations of polarization singularities in the near fields of metaphotonic structures are attracting surging attention. Together with direct observations of polarization singularities in near field assisted by powerful instruments like NSOMs, the way to experimental realizations of nanoscale-controlled light-matter interaction has been significantly  broadened through concepts of singular optics.

\section{Conclusion and Outlook}

We have reviewed the recent theoretical and experimental studies of polarization singularities that are present in both far and near fields generated by nanoscale metaphotonic structures. First, we have discussed the polarization singularities in the far field of finite scattering objects such as a dielectric sphere, which are
the fundamental building blocks of metaphotonic structures governed by the Mie resonances and the Mie theory. In such problems,  intensity null points play an important role, being linked to high-$Q$ resonant modes excited in the particles. Recent studies revealed that the intensity null points carry also vector polarization singularities in the far field. They could be transformed into more generic polarization singularities, the $C$-points. The Poincer\'e-Hopf theorem can be employed to describe such singularities affecting a global charge conservation on the whole far-field momentum sphere.

Next, we have proceeded with the discussion of far-field polarization singularities in one of the most widely discussed metaphotonic structures: a two-dimensional periodic photonic structure, which covers the cases of photonic crystals and metasurfaces.   Starting from the first evidence of polarization singularities in the momentum space of a photonic crystal slab, we have discussed other discoveries of polarization singularities in the far field defined on photonic bands of two-dimensional periodic structures. We have demonstrated that their origin can be explained by the multipolar expansion method implemented for isolated particles. Among the momentum-space polarization singularities, vector polarization singularities, observed as intensity null points in the far field, are known to be related to the concept of bound states in the continuum. Subsequently, we have described the method utilizing these vector polarization singularities to generate phase singularities in the far field. Both passive vortex modulation and active vortex lasing action have been demonstrated. Extending the discussion on polarization singularities to scenarios with photonic band degeneracies, we have reviewed the studies on non-Hermitian degeneracies spawning from the Dirac points, which could serve as a direct platform for establishing the correspondences between polarization singularities and geometric phases.

Polarization singularities also play an important role in the near fields of metaphotonic structures dominating their interactions with matter such as quantum emitters. Control over polarization singularities may enhance the coupling efficiency. We have discussed briefly related theoretical discoveries and propositions. Near-field microscopy based on apertures and tips was invented and developed earlier to study the near fields of structures, so that the developed phase- and polarization-resolved near-field microscopy are great tools for experimental observation and investigation of near-field polarization singularities.

We believe that discoveries of polarization singularities in metaphotonics have deepened our understanding of optical phenomena at the nanoscale. More comprehensive insights can be delivered by blending the concepts from singular optics, topological photonics, and non-Hermitian physics. Such studies may be extended further to time-varying metaphotonic systems such as pulsed structures and moving or rotating bodies~\cite{zdagkas2019singularities}. There may also be a deeper link between the polarization singularities in metaphotonic structures and spontaneous half-vortices in exciton-polaritons \cite{rubo2007half, lagoudakis2009observation, manni2012dissociation}. Other optical singularities also require further research in metaphotonics in both near field and far field, such as caustics \cite{zannotti2020shaping}, optical vortices being singularities of phase~\cite{schouten2003creation, schouten2003light, schouten2004optical, alu2007higher, gorodetski2008observation, yang2009miniature, chen2010experimental, kim2010synthesis, gorodetski2010plasmonic, cho2012coupling, shitrit2012spin, ku2013deterministic, gorodetski2013generating, de2014optical, tsai2014selective, ku2015direct, chen2015creating, spektor2015metafocusing, spektor2017revealing, tan2017controlling, chen2017valley, ostrovsky2018nanoscale, zhu2018topological, yang2018visualization, tsai2019twisted}, vortices of optical currents~\cite{bashevoy2005optical, tribelsky2006anomalous, lukyanchuk2007peculiarities, ahn2012electromagnetic, boriskina2012molding, boriskina2013plasmonics, boriskina2013plasmonic, rahmani2014beyond, tong2014local, yuan2019plasmonics, berry2019roadmap}, and Riemann-Silberstein vortices being singularities in the electromagnetic field independent of the gauge transformation~\cite{bialynicki2003vortex}. We believe these future efforts will broaden significantly the horizons of nanophotonics and offer new perspectives for metaphotonics applications. In particular, further progress in metadevices can be achieve by incorporating optical singularities of various types, leading to novel applications in the near future being definitely beyond our current imagination.

\section{Acknowledgements}

Y. K. acknowledges useful discussions with Prof. Anton Desyatnikov and Prof. Qinghai Song, as well as a support from the Strategic Fund of the Australian National University and the Australian Research Council (grants DP200101168 and DP210101292). Wei L. acknowledges collaborations with Prof. Yuntian Chen, Dr. Weijin Chen, and Mr. Qingdong Yang, and the support from the National Natural Science Foundation of China (grants 11774063 and 11874426).

\bibliography{main.new,References_scattering3}

\end{document}